\documentclass[prd, twocolumn, nofootinbib, floatfix]{revtex4-1}

\usepackage{amsmath}
\usepackage{graphicx}
\usepackage{dcolumn}
\usepackage{bm}
\usepackage{epsfig}
\usepackage{amssymb,latexsym,mathrsfs}
\usepackage{graphicx}
\usepackage{color}
\usepackage{hyperref}

\newcommand{\be}{\begin{equation}}
\newcommand{\ee}{\end{equation}}
\newcommand{\bs}{\begin{split}} 
\newcommand{\bea}{\begin{eqnarray}}
\newcommand{\eea}{\end{eqnarray}}

\newcommand{\om}{\Omega_m}
\newcommand{\ode}{\Omega_{de}}
\newcommand{\onu}{\omega_\nu}
\newcommand{\ocd}{\omega_c}
\newcommand{\ob}{\omega_b}

\usepackage{aas_macros}

\begin{document}

\title{By Dawn's Early Light: CMB Polarization Impact on 
Cosmological Constraints} 
\author{Sudeep Das$^{1}$ \& Eric V.\ Linder$^{1,2}$} 
\affiliation{$^1$Berkeley Center for Cosmological Physics \& Berkeley Lab, 
University of California, Berkeley, CA 94720, USA \\ 
$^2$Institute for the Early Universe WCU, Ewha Womans University, 
Seoul, Korea}

\begin{abstract}
Cosmic microwave background polarization encodes information not only 
on the early universe but also dark energy, neutrino mass, and gravity 
in the late universe through CMB lensing.  Ground based surveys such as 
ACTpol, PolarBear,  SPTpol significantly complement cosmological constraints from 
the Planck satellite, strengthening the CMB dark energy figure of merit  and neutrino mass 
constraints by factors of 3-4.  This changes the dark energy probe landscape.  
We evaluate the state of knowledge in 2017 from ongoing experiments 
including dark energy surveys (supernovae, weak lensing, galaxy clustering), 
fitting for dynamical dark energy, neutrino mass, and a modified 
gravitational growth index.  Adding a modest strong lensing time delay 
survey improves those dark energy constraints by a further 32\%, and 
an enhanced low redshift supernova program improves them by 26\%. 
\end{abstract}

\date{\today} 

\maketitle

\section{Introduction} 

The cosmic microwave background (CMB) radiation temperature and 
polarization power spectra have provided a cornucopia of cosmological 
information, on both the early and late universe \cite{wmap7, arbar,quad, bicep, act, spt}. 
In the next five years considerably more detailed information will 
delivered by the Planck satellite \cite{planck} and ground based, 
high resolution polarization experiments such as ACTpol \cite{actpol}, 
POLAR \cite{polar}, PolarBear \cite{polarbear} and SPTpol \cite{sptpol}.

Such high resolution experiments carry substantial information not 
only on the primordial perturbations but on the matter density power 
spectrum at redshifts $z\approx1-5$.  The matter perturbations 
gravitationally lens the primordial CMB and from the reconstructed 
deflection field one can extract the cosmological parameters that 
impact the matter power spectrum.  These include properties of dark 
energy, including the time variation of its equation of state, the 
sum of neutrino masses, and the gravitational growth index to test 
general relativity. 

In this article we focus on the role of such near term CMB polarization 
information in constraining these cosmological parameters simultaneously, 
since all of them affect the CMB lensing.  We will also bring in other 
dark energy survey information expected within the five year horizon to 
complement the CMB, to give an overall reasonable assessment of the 
constraints in 2017.  We emphasize that the CMB is the main focus of this 
article and we take a fiducial combination of dark energy surveys with 
no intent to study all possible permutations. 

Section~\ref{sec:cmb} gives a brief review of CMB lensing and describes 
the near term CMB data sets.  In Sec.~\ref{sec:other} we present the 
cosmological parameter set and the Fisher matrix constraints from the 
CMB, then adding in various dark energy probes.  Leverage from possible 
further near term data is discussed in Sec.~\ref{sec:further}.

\section{CMB Polarization and Lensing} \label{sec:cmb} 
Recent high resolution instruments have enabled us for the first time to robustly detect and utilize minute distortions in the CMB due to its interactions with the inhomogeneities in the dark matter distribution in the universe, and with the hot gas in galaxy clusters. One unique upshot has been the recent CMB-only detection and characterization of the gravitational lensing signal, marking the beginnings of an exciting new field \cite{actdetection, sptdetection}. The CMB lensing effect is conceptually simple: CMB photons propagating through the universe get deflected by inhomogeneities in the large scale distribution of matter, so that the image of the CMB at our end is a distorted version of the image at the last scattering surface where the CMB was formed \cite{lewis_and_challinor_1998}. The rms deflection of the CMB photon is about 3 arcminutes on the sky and is caused by all the structure between us and the last scattering surface with the main contribution coming from the redshift range  $z\approx 1-5$. The deflections are coherent over degree angular scales, and coherently distort arcminute scale anisotropies in the CMB. This leads to smearing of the peaks in the CMB temperature power spectrum and also the generation of small scale B mode polarization from the lensing of E-modes.  A unique aspect of lensing is that it induces a specific form of non-Gaussianity in the CMB, which can then be used to reconstruct the projected dark matter potential field. This projected lensing potential depends both on the geometry and growth of structure. Dark energy affects both geometry and growth, and massive neutrinos affect growth of structure by suppressing the amplitude of matter fluctuations on scales smaller than their free streaming length. These effects make CMB lensing one of the most promising probes of neutrino mass sum and dark energy behavior \cite{sherwin_etal_2012, dePutter_2009}. \par

The next generation of polarization enabled high resolution CMB experiments will be primarily CMB lensing machines.  Since both temperature and polarization  are lensed by the same large scale deflection field, the inclusion 
of polarization improves the fidelity of the lensing reconstruction. Also, foregrounds are expected to be much 
lower in polarization, enabling us to use more modes to perform CMB lensing reconstruction than temperature alone. (Although in this article we use a conservative limit of  $\ell_{\rm {max}} = 3000$ for both temperature and polarization modes, the expectation is that we should be able to use modes up to $\sim 5000$ in polarization.)  For the projections made here, we consider an amalgam of high resolution CMB polarization data from ACTpol, PolarBear, and SPTpol.  We assume a total of 10,000 deg$^2$ covered at the depth of 5 $\mu$K-arcmin in temperature and 7 $\mu$K-arcmin in polarization. Although the resolutions of these experiments vary, we assume a uniform FWHM 
of 1 arcmin for the entire combined dataset. This is justified because, at the depth considered here, and with the range of CMB multipoles used for the lensing reconstruction, the projections are largely insensitive to the beam width as long as it is below 4 arcmin.  \par
For the CMB-only projections (especially those not including lensing information) large degeneracies exist in several beyond-$\Lambda$CDM parameters. For the projections in such cases (e.g.\ $w_0$-$w_a$ contour for Planck without lensing), the large Fisher error ellipses shown here should be viewed as qualitative. 
 
\section{Cosmological Constraints} \label{sec:other} 

\subsection{CMB Surveys} \label{sec:cmbs} 

The usual set of cosmological parameters describing the vanilla model 
is the dimensionless physical baryon density $\ob=\Omega_b h^2$, cold 
dark matter density $\ocd=\Omega_c h^2$, dark energy density $\ode$, 
spectral index of scalar perturbations $n_s$, optical depth $\tau$, 
and amplitude of mass perturbations $\sigma_8$.  The dimensionless 
Hubble parameter $h$ is defined implicitly through the summation of the 
energy densities to the critical value; we restrict to a spatially flat 
universe as a theoretical prior.  

We enlarge this set of parameters by considering the main physical 
quantities that would impact the matter power spectrum and hence CMB 
lensing.  Dark energy other than a cosmological constant is likely to 
have a dynamical equation of state $w(a)$, with $a$ the cosmic scale 
factor.  The form $w(a)=w_0+w_a(1-a)$ was fit from exact solutions of 
scalar field theories \cite{lin03} and derived as a calibration relation 
for observables accurate at the 0.1\% level \cite{calib}, so we include 
the two parameters $w_0$, the present dark energy equation of state, and 
$w_a$, a measure of its time variation.  

Neutrinos have mass, and this 
affects the matter power spectrum, so we include the dimensionless neutrino 
energy density $\onu=\Omega_\nu h^2$ as a parameter.  Note that the sum 
of the neutrino masses $\sum m_\nu=94\,\onu\,$eV.  Finally, although within 
general relativity the background densities determine the growth rate of 
matter density perturbations, in other theories the strength of gravity, 
the effective Newton's constant, has an additional impact.  To test for 
deviations from general relativity (GR) we include the gravitational growth 
index $\gamma$ \cite{lin05}, a single parameter shown to describe the 
growth deviations in several models of scale-independent modified gravity 
at the 0.1\% level \cite{lincahn}.  These 10 parameters are simultaneously 
fit; as we will see, accounting for only a subset of the additional 
parameters can significantly overestimate the constraining power, or bias the 
results in the case of an improper fiducial (e.g.\ assuming zero neutrino 
mass).  Fiducial values are given in Table~\ref{tab:pol10k}.

\begin{table*}[!htb]
\begin{tabular}{l|cccccccccc}
&$\ob$ &$\ocd$ &$\onu$&$\ode$&$n_s$&$\tau$&$\sigma_8$&$w_0$&$w_a$&$\gamma$\\ 
\hline 
Fiducial& 0.02258& 0.1093& 0.001596& 0.734& 0.963& 0.086& $\ $0.8$\ $& -1& 0& $\ $0.55$\ $\\ 
$\sigma$(Planck)& 0.000137& 0.00117& 0.00175& 0.124& 0.00337& 0.00426& d& 1.10& 2.48& d\\ 
$\sigma$(Planck+10k)$\ $ & $\ $0.0000492$\ $ & $\ $0.000682$\ $ & $\ $0.000666$\ $& $\ $0.042$\ $ & $\ $0.00207$\ $& $\ $0.00297$\ $ & d& $\ $0.305$\ $& $\ $0.642$\ $ & d\\ 
Gain& 2.78& 1.72& 2.63& 2.95& 1.63& 1.43& d& 3.61& 3.86& d\\ 
\end{tabular}
\caption{$1\sigma$ constraints from forthcoming CMB data on cosmological 
parameters, from the Planck satellite and from Planck plus 10,000 deg$^2$ 
of high resolution, ground based CMB polarization data.  Gain shows the 
improvement due to adding the ground data.  The degeneracy (``d'') between 
the parameters $\sigma_8$ and $\gamma$ is not well broken.  The derived 
fiducial $h=0.708$ and $\sum m_\nu=0.15\,$eV. 
}
\label{tab:pol10k}
\end{table*}

\begin{figure}[htbp!]
\includegraphics[width=\columnwidth]{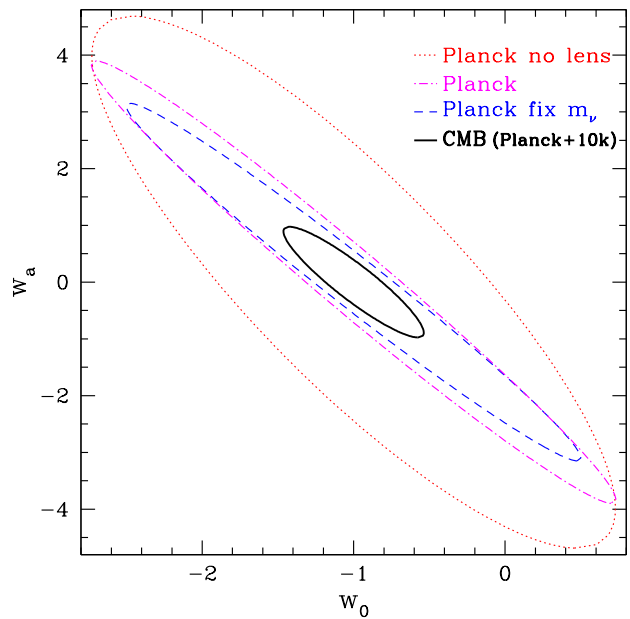}
\caption{68\% confidence contours on the dark energy equation of state 
parameters $w_0$ and $w_a$, marginalized over all remaining parameters, 
for various cases of CMB data.  All data uses both CMB temperature and 
polarization power spectra, but in the ``no lens'' case does not use 
the information in the CMB lensing deflection field.  The ``fix $m_\nu$'' 
case fixes the sum of neutrino masses to the fiducial.  Ground based, 
high resolution data is included in the $+10k$ case and delivers a 
substantial improvement in constraints. 
}
\label{fig:cmbgross}
\end{figure}

\begin{figure}[htbp!]
\includegraphics[width=\columnwidth]{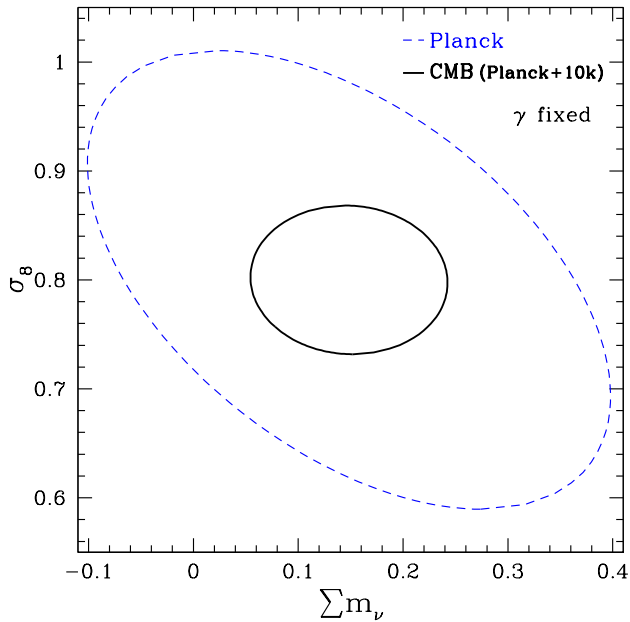}
\caption{68\% confidence contours on the 
parameters $\sigma_8$ and $\sum m_\nu$, marginalized over all remaining parameters, 
except the growth index $\gamma$ which is  fixed at its GR value, 
for various cases of CMB data.  
}
\label{fig:sig8mnu}
\end{figure}

Figure~\ref{fig:cmbgross} illustrates the constraints in the dark energy 
equation of state plane for various cases using CMB data.  The most 
important lesson is that including CMB lensing information gives dramatic 
improvement in the constraints.  The second lesson is that, as just stated, 
fixing parameters such as neutrino mass, rather than simultaneously fitting 
for them, can underestimate the errors.  Finally, we see that near term, high 
resolution ground based CMB polarization surveys generate 
significantly tighter constraints.  This gain applies to all parameters, 
not just the dark energy equation of state ones, as Table~\ref{tab:pol10k} 
shows.  In particular, the gain in the constraint on neutrino mass (assuming 
GR, here only, to break the $\gamma-\sigma_8$ degeneracy)  is significant, as shown in Fig.~\ref{fig:sig8mnu}.

\subsection{Dark Energy Surveys} \label{sec:des} 

We now consider the addition of dark energy survey probes, treating as 
fiducial CMB data the combination of Planck plus 10,000 deg$^2$ of ground 
based, high resolution data, accounting for CMB lensing and fitting all 
parameters simultaneously.  These other probes may bring in further, 
nuisance parameters, which we also marginalize over.  Our guiding principle 
is to consider surveys essentially underway, giving a quantitative level 
of results neither optimistic nor pessimistic, but reasonably expected 
within 5 years. 

Distances from the Type Ia supernova (SN) magnitude-redshift measurements 
play an important role in constraining the dark energy equation of state. 
We adopt a combination of projected data of 150 SN at $z=0.03-0.08$ from 
the Nearby Supernova Factory \cite{snf}, 1000 SN at $z=0.1-1$ from the 
Dark Energy Survey (DES \cite{des}) and other SN surveys, and 42 SN at 
$z=1-1.7$ from Hubble Space Telescope surveys.  These numbers are not 
to be interpreted as the total numbers expected (for example DES should 
get several thousand SN) but rather the number in line with the level of 
systematic errors adopted.  Systematic uncertainties of $0.02(1+z)$ mag 
per 0.1 redshift bin are added in quadrature with individual intrinsic 
dispersions of 0.13 mag.  In Sec.~\ref{sec:further} we consider the 
possibility of enhanced data, but we adopt this as a canonical, reasonable 
data set. 

Weak gravitational lensing (WL) probes both distances and growth and is key 
in breaking the degeneracy between $\sigma_8$ and $\gamma$.  We model the 
projected data on DES, with 5000 deg$^2$ of galaxy shears measured at a 
density of $n_g=12/$arcmin$^2$ and a median redshift depth $z_{\rm med}=0.7$. 
Systematic uncertainties with priors in photometric redshift bias of $10^{-3}$, 
in photo-z scatter of $10^{-3}$, multiplicative shear of $10^{-2}$, and 
additive shear of $10^{-5}$ are used.  See \cite{berwl} for more details 
on the weak lensing computational code that includes the whole set of 
our cosmological parameters affecting growth. 

Galaxy clustering can be used as a distance (and Hubble parameter) 
estimator through the baryon acoustic oscillation (BAO) scale of 
wiggles in the matter power spectrum, and a probe of growth through the 
full shape and amplitude of the power spectrum.  Angular anisotropies 
due to peculiar motions, called redshift space distortions, carry 
information on the growth rate and constrain both the background expansion 
and the gravitational growth index $\gamma$.  We model the projected 
power spectrum data (PK) 
on the ongoing Baryon Oscillation Spectroscopic Survey (BOSS \cite{boss}), 
using the power spectrum out to $k_{\rm max}=0.125\,h$/Mpc and allowing 
for galaxy bias fit parameters in each of the redshift bins at $z=0.35$, 0.6. 

\begin{figure}[htbp!]
\includegraphics[width=\columnwidth]{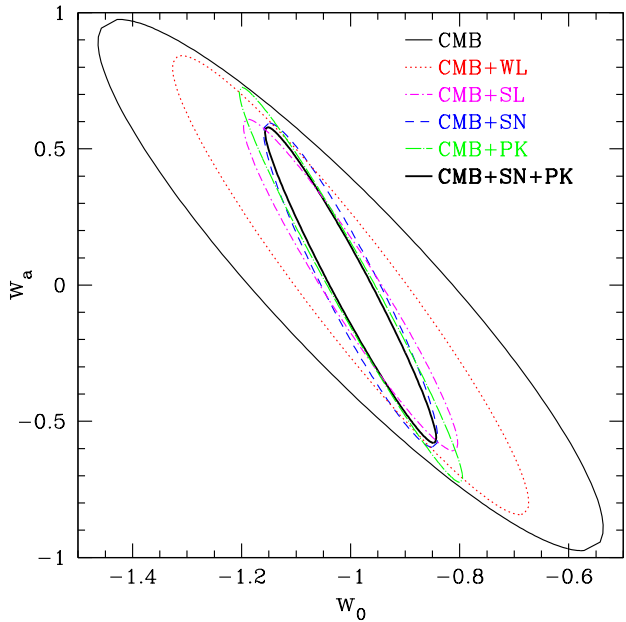}
\caption{68\% confidence contours of $w_0$ and $w_a$ for combinations 
of other surveys with CMB data.  Supernova (SN) surveys are particularly 
complementary for dark energy parameters, but galaxy clustering (PK) 
is constraining in other parameters such as $\gamma$.  The fiducial 
combination of SN+PK surveys with CMB data improves the area uncertainty 
(figure of merit) by a factor 8.1 over CMB alone. 
}
\label{fig:compareall}
\end{figure}

Figure~\ref{fig:compareall} illustrates the impact of complementarity 
of the dark energy surveys, one by one, with the CMB polarization data.  
The differing orientation of the confidence contours shows further 
complementarity will come from combining several probes.  Indeed the 
combination CMB+SN+PK improves the contour area figure of merit 
(FOM=$1/\sqrt{\det COV[X,Y]}$, where $COV$ is the covariance matrix of 
parameters $X$, $Y$, marginalized over all others) by 44\% over CMB+SN 
and 42\% over CMB+PK.

\begin{figure}[htbp!]
\includegraphics[width=\columnwidth]{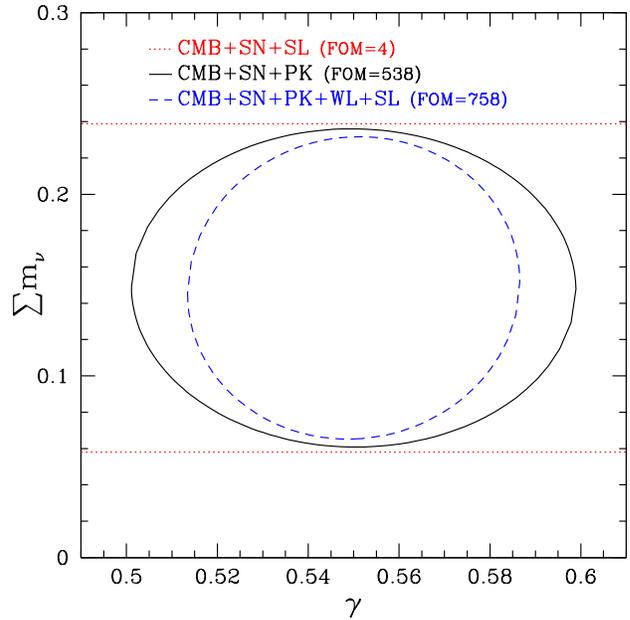}
\caption{68\% confidence contours on the sum of neutrino masses 
$\sum m_\nu$ and  gravitational growth index $\gamma$, marginalized over 
all remaining parameters, for combinations of other surveys with CMB data.  
PK breaks the growth covariance, determining $\gamma$; CMB data provides 
nearly all of the constraint on neutrino mass, to an uncertainty of 0.058 eV. 
Inverse contour areas are indicated by the values FOM. 
}
\label{fig:mnugam}
\end{figure}

However gains are simultaneously made in the other parameters; in particular 
PK allows separate and precise determination of $\sigma_8$ and $\gamma$.  
Figure~\ref{fig:mnugam} shows the uncertainties in the $\gamma$--$\sum m_\nu$ 
plane.  Note how the CMB is capable of determining the neutrino masses, 
but without separate growth information from PK the degeneracy between 
$\sigma_8$ and $\gamma$ prevents reasonable determination of $\gamma$.

From Fig.~\ref{fig:compareall} we see there is good complementarity between 
each of PK, SN, and strong lensing time delays (SL, discussed in 
Sec.~\ref{sec:sltime}).  WL is also complementary but has less leverage 
at the level of systematics taken for DES.  If the systematics can be 
overcome, in DES or a further future experiment, then WL will be more 
powerful.  On the other hand, the complementarity of the other probes 
indicates that if WL falters, then this will not badly harm our knowledge 
of the cosmological parameters.  In particular, in the presence of PK data 
then SL can meet or exceed the contributions of WL.  The one exception is 
the gravitational growth index $\gamma$. 

The parameter constraints from our fiducial projections for data within 
the next five years, representing the combination CMB+SN+PK, appear 
in Table~\ref{tab:combo}.  We also present the results including WL, and 
SL as in Sec.~\ref{sec:sltime}.

\begin{table*}[!htb]
\begin{tabular}{l|ccccccccccc}
&$\ 10^5 \ob$ &$10^4 \ocd$ &$10^4 \onu$&$\ode$&$n_s$&$\sigma_8$&$w_0$&$w_a$&$\gamma$&FOM$w$ & FOM$\nu$\\ 
\hline 
$\sigma$(CMB+SN+PK)$\ $& $\ $4.76 & 6.47 & 6.21 & 0.00507 & 0.00200 & 0.0110 & 0.103 & 0.382 & 0.0322 & 103 & 538 \\ 
$\sigma$(CMB+SN+PK+WL)$\ $ & $\ $4.71 & 5.85 & 5.97 & 0.00470 & 0.00192 & 0.00934 & 0.0927 & 0.339 & 0.0256 & 120 & 704 \\ 
$\sigma$(CMB+SN+PK+SL)$\ $ & $\ $4.74 & 6.03 & 6.12 & 0.00414 & 0.00195 & 0.0107 & 0.0801 & 0.292 & 0.0319 & 135 & 551 \\ 
$\sigma$(CMB+SN+PK+WL+SL)$\ $ & $\ $4.70 & 5.63 & 5.89 & 0.00403 & 0.00189 & 0.00808 & 0.0774 & 0.280 & 0.0241 & 147 & 758 \\ 
\end{tabular}
\caption{$1\sigma$ constraints from forthcoming data on cosmological 
parameters, from various combination of probes on the 2017 timescale. 
We consider the first combination, CMB+SN+PK, to be the baseline and also 
show various extensions. 
Here FOM$w$=FOM[$w_0,w_a$] is the dark energy figure of merit and 
FOM$\nu$=FOM[$\sum m_\nu,\gamma$] is a growth figure of merit. 
}
\label{tab:combo}
\end{table*}

\section{Additional Information} \label{sec:further} 

\subsection{Time Delay Distances} \label{sec:sltime} 

Geometric measures of distance ratios through strong gravitational 
lensing time delays (e.g.\ of quasars lensed by galaxies) have especially 
good complementarity with SN \cite{linsltime}.  While this technique, 
which requires both imaging and spectroscopy for accurate lens mass 
modeling, is not a canonical part of any of the surveys we have discussed, 
individual groups are pursuing time delay surveys 
\cite{fadely,treu,cosmograil}.  

Here we take a 
modest version of the survey proposed in \cite{linsltime}, requiring 
detailed measurement of a total of 38 time delay systems over the 
range $z=0.1$--0.6, each of 5\% accuracy in the time delay distance 
(equivalent to 2\% accuracy per 0.1 bin in redshift).  To obtain the 
high resolution needed to map accurately the multiple images and 
create a robust lens mass model (see, e.g., \cite{oguri,suyu08}), 
Hubble Space Telescope 
imaging is required.  As a rough estimate, each time delay system 
requires $\sim$6 orbits \cite{suyuprivate}, so the entire program (not 
counting the systems already investigated) would take $\sim$230 orbits.  
A restricted range of $z=0.3$--0.6 only loses 2\% on FOM($w_0,w_a$) 
and FOM($m_\nu,\gamma$) while reducing the time to $\sim$150 orbits. 

\begin{figure}[htbp!]
\includegraphics[width=\columnwidth]{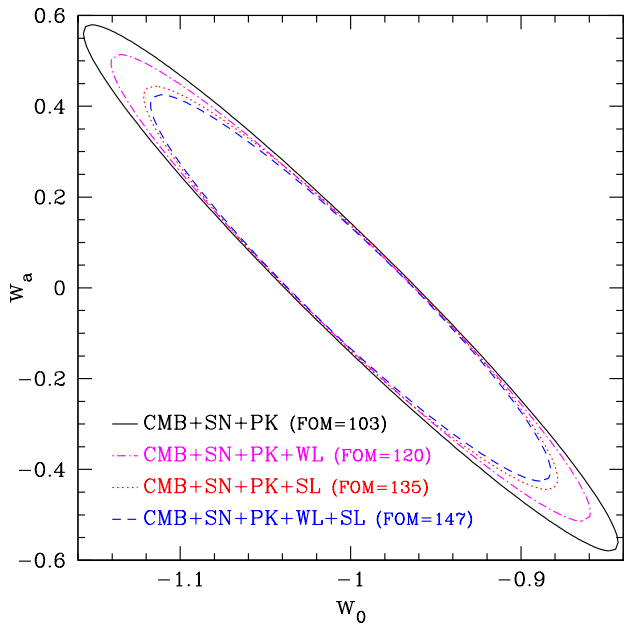}
\caption{68\% confidence contours of $w_0$ and $w_a$ for combination of 
strong lensing time delay data (SL) with other surveys.  Even relative 
to the fiducial CMB+SN+PK, the SL data has further complementarity, 
improving the FOM by 32\%, a stronger gain than from WL. 
}
\label{fig:comparesl}
\end{figure}

The cosmological impact of such a strong lensing (SL) survey is dramatic.  
The complementarity with SN (and other probes) is evident in 
Fig.~\ref{fig:compareall}.  
Added to CMB+SN, SL provides a gain of 52\% in dark energy equation of 
state FOM.  Even added to the fiducial CMB+SN+PK, the constraints improve 
further, by 32\%, as illustrated in Fig.~\ref{fig:comparesl}.  The gain 
over the fiducial survey combination is 18\% in $\om$.

While the time delay probe is not as mature at the moment as other 
techniques, its potential leverage makes it worth more detailed consideration 
for allocating observing time over the next five years.  If systematic 
uncertainties can be controlled to better than 5\% per time delay 
system, then the observing time required can be reduced.  That is, 
we assumed 2\% distance per bin; if systematics can be reduced to 
3.5\% per system then only half as many systems need to be used. 
Recent work \cite{suyu12} also offers hope for using simpler two image 
systems.

\subsection{Enhanced Surveys} 

We now consider the impact on cosmological parameter estimation for 
some specific enhancements to the surveys.  This is not an optimization 
exercise but rather an exploration of possible further leverage. 

\begin{figure}[htbp!]
\includegraphics[width=\columnwidth]{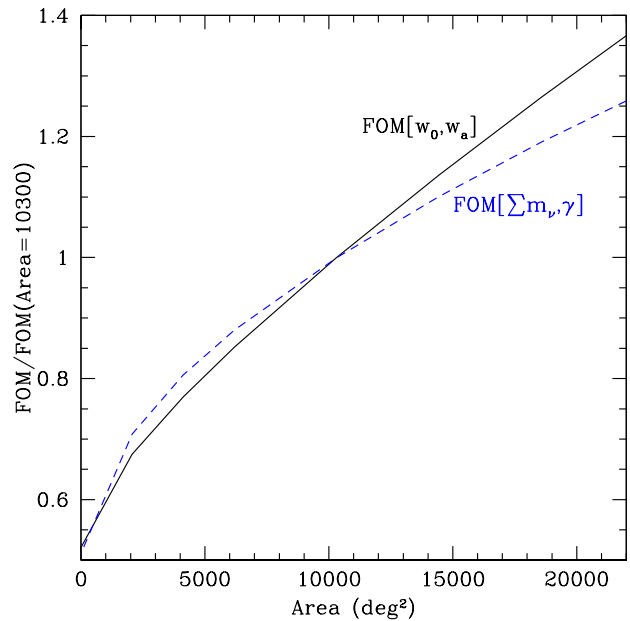}
\caption{For the baseline survey combination CMB+SN+PK the dark energy 
and growth figures of merit increase as shown with increasing area of the ground based CMB survey.  
}
\label{fig:cmbarea}
\end{figure}

First we look at the CMB ground polarization data.  While observations 
from the South Pole are restricted to $\sim4000$ deg$^2$ of sky, 
mid latitude sites in both hemispheres could do larger areas.  We thus 
consider how the cosmology constraints vary for areas other than our fiducial 
10,000 deg$^2$ (at the same depth).  For areas where Planck and ground-based surveys overlap, we 
add their inverse noises in quadrature. Fig.~\ref{fig:cmbarea}  illustrates the dependence
of the dark energy and neutrino-gravity (growth) figures of merit on the area of the ground based survey.

\begin{figure}[htbp!]
\includegraphics[width=\columnwidth]{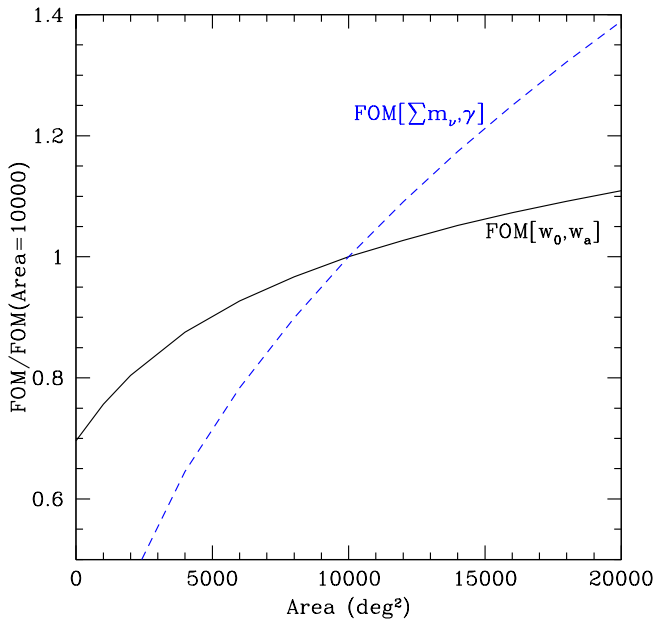}
\caption{For the baseline survey combination CMB+SN+PK the dark energy 
and growth figures of merit increase with increasing area of the galaxy 
clustering (PK) survey.  The growth figure of merit, describing our 
knowledge of neutrino mass and ability to test Einstein gravity, improves 
more quickly with area. 
}
\label{fig:pkarea}
\end{figure}

Another possibility is to increase the area of the spectroscopic 
galaxy survey beyond the 10000 deg$^2$ of our fiducial.  Note that a 
greater area survey would likely extend beyond our 5 year horizon. 
Figure~\ref{fig:pkarea} shows that the growth figure of merit capturing 
the sum of neutrino masses $\sum m_\nu$ and the gravity growth index 
$\gamma$ continues to increase significantly for larger surveys, while 
the dark energy figure of merit is closer to saturation.

For the supernova dark energy probe, a particularly useful enhancement 
on the 5 year time scale is a larger nearby supernova survey with 
tightly controlled systematics.  Our fiducial involved 150 supernovae 
with a systematics floor of 0.021 mag.  Considerable effort is ongoing 
at improving supernova systematics with highly calibrated spectrophotometry 
{\`a} la the Nearby Supernova Factory \cite{snf} and near infrared 
observations {\`a} la the Carnegie Supernova Project \cite{csp}, both 
offering promising leads.  We therefore consider an enhanced program 
with 300 supernovae in $z=0.03-0.08$ with a systematics floor over this 
range of 0.008 mag.  This seems within the realm of practicality for 
our 5 year timescale. 

Such a supernova program yields significant improvements.  For the 
combination CMB+SN+PK the $w_0$--$w_a$ FOM rises by 26\%, from 103 to 
129.

\subsection{Fixing Parameters} \label{sec:fix} 

Our approach has been to include all the main cosmological parameters 
affecting the matter power spectrum: dynamical dark energy ($w_0$, $w_a$), 
neutrino mass ($m_\nu$), and the gravitational deviations ($\gamma$).  
Ignoring any of these by fixing a parameter will produce higher figures of 
merit, at the price of neglecting possibly relevant physics.  This seems 
particularly undesirable in the case of neutrino mass, where we know the 
physics exists yet we do not know the parameter value. 

\begin{figure}[htbp!]
\includegraphics[width=\columnwidth]{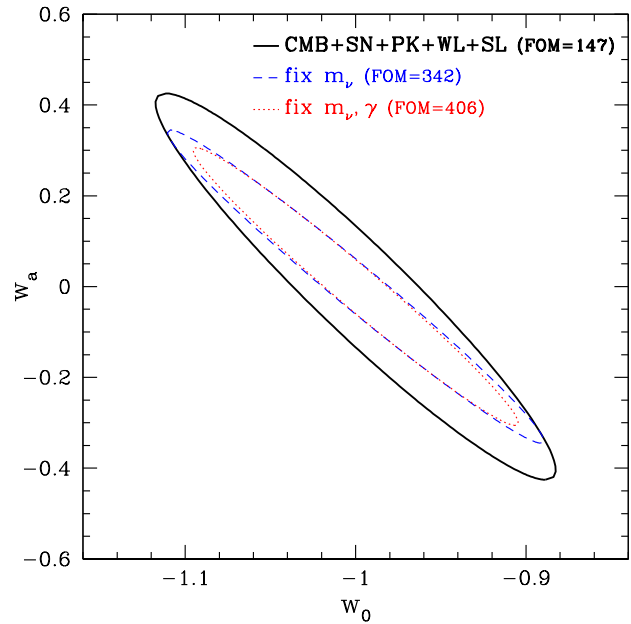}
\caption{68\% confidence contours of $w_0$ and $w_a$ significantly 
shrink if we neglect, or assume perfect knowledge of, neutrino mass 
and gravity.  Such assumptions are not generally justified. 
}
\label{fig:fixmnu}
\end{figure}

Figure~\ref{fig:fixmnu} shows the effect of fixing $\onu$ 
(i.e.\ $\sum m_\nu$) on the dark energy estimation.  The major degeneracy 
that is artificially broken by ad hoc fixing is with the pivot value of 
the dark energy equation of state, $w_p$, which decorrelates this value 
from the time variation $w_a$.  Neglecting neutrino mass or assuming its 
value makes FOM appear to be 2.3 times higher than it should be.  Strong 
lensing time delays are particularly powerful in the fixed $m_\nu$ case, 
increasing the FOM by 76\% rather than the previous 32\% of 
Fig.~\ref{fig:comparesl}.

If we disallow deviations from general relativity, fixing $\gamma$, 
then the main effect is breaking the degeneracy with $\sigma_8$.  In 
such a case galaxy clustering (or weak lensing) is no longer needed to 
determine $\sigma_8$ 
as the CMB and its lensing do a good job (for example determining it 
to 0.01 for CMB+SN, and to 0.007 for CMB+SN+SL).  For the combination 
with galaxy clustering, fixing $\gamma$ leads to $\sigma_8$ constrained 
a factor 1.7 
more tightly than with $\gamma$ fit.  The effect on FOM$w$ is much more 
modest, just 8\%.  Fixing both $\onu$ and $\gamma$ basically convolves 
the effects of each.  For the FOM using all probes the total induced 
tightening is a factor 2.8; i.e.\ FOM$w$ would be 406 in the commonly used 
parameter space ignoring $m_\nu$ and $\gamma$, rather than 147.  
We emphasize that fixing parameters, apart from restricting 
physics perhaps unjustifiably, opens the result to bias if the wrong 
parameter values are used.

\section{Conclusions} \label{sec:concl} 

The origin of cosmic acceleration is a fundamental mystery of physics, 
and the subject of active and varied observational efforts.  We find and 
quantify that the newly maturing probe of CMB lensing can play a 
significant role. 
Within the next 5 years maps of the CMB lensing deflection field over wide 
areas will be obtained, considerably strengthening our cosmological 
knowledge.  These experiments are underway or imminent, and combination 
with ongoing or near term dark energy surveys will provide a large step 
forward in evaluating dark energy, neutrino, gravity, and primordial 
perturbation properties. 

The high resolution, low noise, ground based CMB polarization experiments 
give powerful leverage.  Indeed, in combination with all other survey 
probes in the commonly used, ``vanilla plus $w_0$, $w_a$'' parameter space 
the result would be a very strong dark energy figure of merit 
FOM[$w_0,w_a$]=406.  However, it is overly restrictive to assume perfect 
knowledge of the sum of neutrino masses and the behavior of gravity, both 
of which affect the growth of structure we are using as a cosmological probe. 
We present results including simultaneous fits for not only the vanilla 
parameters but dark energy dynamics $w_0$ and $w_a$, gravity $\gamma$, 
and sum of neutrino masses $\sum m_\nu$.

Concentrating on data expected to arrive within the next 5 years, we 
examine constraints from realistic near term experiments measuring CMB 
satellite and ground based polarization, supernova distances, and the 
galaxy power spectrum, as the baseline combination of surveys.  The 
complementarity between the different probes is highlighted and quantified.  
In addition to the cosmological parameter estimation and figures of merit in 
dark energy and gravity--neutrino planes, we analyze the dependence of the 
results on the sky area of the CMB space/ground overlap and of galaxy surveys. 
We further study the impact of weak lensing data.  

Interestingly, we find that 
substantial further leverage can be obtained in supplementing the 
canonical program with a modest strong lensing time delay survey, 
requiring of order $\sim$200 orbits on the Hubble Space Telescope, 
and a tightly systematics controlled nearby supernova program.  
This overall combination of experiments would deliver knowledge of 
the dark energy time variation $w_a$ to 0.25 and of the sum of the 
neutrino masses to 0.055 eV with data taken by $\sim$2017. 

CMB lensing thus acts a powerful component of a dark energy and cosmology 
survey program, enabling us to close in chasing down cosmic acceleration.  
For further improvements between 2017 and when the next generation of 
cosmic volume surveys deliver data, promising paths forward include 
the  use of cross-correlations between probes, the development of new probes (just as CMB lensing and strong lensing 
grew viable), reduction of weak lensing systematics, and the ability 
to accurately use smaller scales ($k_{max}>0.125\,h$/Mpc) in the galaxy 
power spectrum.  

\acknowledgments

This work has been supported in part by the Director, 
Office of Science, Office of High Energy Physics, of the U.S.\ Department 
of Energy under Contract No.\ DE-AC02-05CH11231 and by World Class 
University grant R32-2009-000-10130-0 through the National Research 
Foundation, Ministry of Education, Science and Technology of Korea.


\end{document}